\documentclass[a4paper,11pt]{article}
\usepackage{pos}

\usepackage{xcolor}

\usepackage{braket}
\usepackage{graphicx}
\usepackage{amsmath}
\usepackage{extarrows}

\newcommand{\be}{\begin{equation}}
\newcommand{\ee}{\end{equation}}
\newcommand{\bea}{\begin{eqnarray}}
\newcommand{\eea}{\end{eqnarray}}

\title{Generalized form factors of the pion and kaon using twisted mass fermions}

\author*[a]{Joseph Delmar}
\author[b,c]{Constantia Alexandrou}
\author[c]{Simone Bacchio}
\author[d]{Ian Clo\"et}
\author[a]{Martha Constantinou}
\author[c]{Giannis Koutsou}

\affiliation[a]{Department of Physics, Temple University,\\ 1925 N. 12th Street, Philadelphia, PA 19122-1801, USA}

\affiliation[b]{Department of Physics, University of Cyprus,\\  P.O. Box 20537,  1678 Nicosia, Cyprus}

\affiliation[c]{Computation-based Science and Technology Research Center,
  The Cyprus Institute,\\ 20 Kavafi Str., Nicosia 2121, Cyprus}

\affiliation[d]{Physics Division, Argonne National Laboratory,\\ Lemont, Illinois 60439, USA}

\emailAdd{jdelmar@temple.edu}

\abstract{We present an update on our lattice calculations of the Mellin moments of PDFs and GPDs for the pion and kaon, using momentum-boosted meson states. In particular, we focus on the calculation of the scalar and tensor local operators, and the vector operator with up to three-covariant derivatives. The corresponding matrix elements allow us to extract the scalar and tensor charges, as well as $\langle x^n \rangle$ with $n=1,2,3$. In addition, we introduce momentum transfer between the initial and final meson state that leads to the scalar, vector, and tensor form factors, and the generalized form factors up to three covariant derivatives. The above results are obtained using two ensembles of maximally twisted mass fermions with clover improvement with two degenerate light, a strange, and a charm quark $(N_f=2+1+1)$ at lattice spacings of 0.093 fm and 0.081 fm. The pion mass of the ensembles is about 260 MeV. We study excited-states effects by analyzing four values of the source-sink time separation (1.12 - 1.67 fm). We also examine discretization and volume effects. The lattice data are renormalized non-perturbatively, and the results are presented in the MS scheme at a scale of 2 GeV.}

\FullConference{%
 The 40th International Symposium on Lattice Field Theory, LATTICE2023
  July 31 - August 4, 2023
  Fermi National Lab
}

\begin{document}

\maketitle

\section{Introduction}
Quantum chromodynamics (QCD) describes the strong interaction, which governs the dynamics of hadrons. The phenomena described by QCD are incredibly complex and typically cannot be described analytically. Instead, simulations of the theory on a discretized space-time lattice can produce distributions which provide insight on hadronic properties such as internal structure. The form factors (FFs) and generalized form factors (GFFs) are related to the internal forces of the hadron and the emergence of their mass~\cite{Polyakov:2018zvc}. There have been a few lattice studies of the quark GFFs of the proton and pion, as well as gluonic contibutions from various hadrons~\cite{Pefkou:2021fni,Hackett:2023nkr,Hackett:2023rif}. Studies of the pion and kaon are of particular interest, as they are the Goldstone bosons that arise from spontaneous chiral symmetry breaking in QCD. 

In these proceedings, we present our results for the quark GFFs of the pion and kaon for two ensembles with different lattice spacings. We consider only results accounting for the connected contributions, as we expect the quark disconnected contributions to be small. We apply parameterizations on the 4-vector momentum transfer squared, $-t$, and examine whether the lattice data favor a monopole or an $n$-pole function. Additionally, we present Mellin moments up to $\langle x^3 \rangle$ and examine SU(3) flavor symmetry-breaking effects in the FFs and GFFs. 

\section{Methodology}

This calculation is carried out on two ensembles of twisted-clover fermions and Iwasaki improved gluons generated by the Extended Twisted Mass Collaboration (ETMC), labelled cA211.30.32 and cB211.25.48 with lattice spacing $a=0.094$ fm and $a=0.079$ fm, respectively~\cite{PhysRevD.104.074520}. Tab.~\ref{tab:lattice_params} provides a summary of the other parameters for these two ensembles. 

\begin{table}[h!]
    \centering
    \begin{tabular}{| c | c | c | c | c | c | c |}
    \hline
    Ensemble  & $\beta$ & $a$ (fm)  & $L^3 \times T$ & $N_f$ & $m_\pi$ (MeV) & $L$ (fm)\\
    \hline
  cA211.30.32 & 1.726 & 0.094 & $32^3 \times 64$ & $2{+}1{+}1$ & 265 & 3.0\\
  cB211.25.48 & 1.778 & 0.079 & $48^3 \times 96$ & $2{+}1{+}1$ & 250 & 3.79\\
    \hline
    \end{tabular}
    \caption{Parameters for the ensembles used in this work.}
    \label{tab:lattice_params}
\end{table}

The decomposition of the matrix elements requires a direction of momentum boost for each order of derivative in the operator. For access to GFFs corresponding to operators with up to three derivatives, we utilize a boosted kinematic frame with final momentum containing the smallest nonzero component ($\pm1$) in each direction for a total of 8 combinations of final momentum. This boosted frame also provides the benefit of extracting a denser range of $-t = \vec{q}^2-(E(p')-E(p))^2$,
where $\vec{q}$ is the momentum transfer in the spatial directions. We also gain access to a higher range of momentum transfer (2.5-3.0 Gev$^2$). Furthermore, we eliminate mixing under renormalization, which is possible due to the kinematic setup. The trade-off for this is a decreased signal-to-noise ratio when compared with a frame utilizing the minimum directions of nonzero momentum. The three-point functions for ensemble cA211.30.32 have been calculated at source-sink time separations, $t_s$, of 1.13, 1.32, 1.50, and 1.69 fm. For ensemble cB211.25.48, we extract three-point functions at $t_s=$1.11, 1.26, 1.42, and 1.58 fm. The statistics for both ensembles is shown in Tab.~\ref{tab:stats}. Calculations on ensemble cA211.30.32 have reached sufficient statistics for high-precision results, while data production on ensemble cB211.25.48 is ongoing.

\begin{table}[h!]
    \centering
    \hspace*{-0.8cm}\begin{tabular}{| c | c | c | c | c | c | c | c|}
    \hline
    Ensemble  & Frame & $\vec{p} (2\pi/L)$  & $t_s/a$ & $t_s/a$ (fm) & confs & src pos. & Total\\
    \hline
  cA211.30.32 & B & $(\pm1,\pm1,\pm1)$ & 12, 14, 16, 18 & 1.13, 1.32, 1.50, 1.69 & 122 & 136 & 132,736\\
  cB211.25.48 & B & $(\pm1,\pm1,\pm1)$ & 14, 16, 18, 20 & 1.11, 1.26, 1.42, 1.58 & 76 & 8 & 4,864\\
    \hline
    \end{tabular}
    \caption{Statistics for the ensembles used in this calculation.}
    \label{tab:stats}
\end{table}

The GFFs we analyze are obtained from the matrix elements of local operators containing one to three covariant derivatives defined as $ O^{\{\mu_1 \cdots \mu_n\}} \equiv \overline{\psi} \gamma^{\{\mu}  D^{\mu_1}\cdots D^{\mu_n\}}\psi $.
The matrix elements decompose to form factors ($n=0)$ and generalized form factors ($n>0$) according to~\cite{Hagler:2009ni}:
\begin{align}
\hskip -0.75cm
    \big\langle M(p') | O^{\{\mu\nu\}} | M(p) \big\rangle &= K\big[2P^{\{\mu}P^{\nu\}}A_{20} + 2\Delta^{\{\mu}\Delta^{\nu\}}B_{20} \big]\\
    \hskip -0.75cm
    \big\langle M(p') | O^{\{\mu\nu\rho\}} | M(p) \big\rangle &= K\big[2iP^{\{\mu}P^{\nu}P^{\rho\}}A_{30} + 2i\Delta^{\{\mu}\Delta^\nu P^{\rho\}}B_{30} \big]\\
    \hskip -0.75cm
    \big\langle M(p') | O^{\{\mu\nu\rho\sigma\}} | M(p) \big\rangle &= K\big[-2P^{\{\mu}P^{\nu}P^{\rho}P^{\sigma\}}A_{40} - 2\Delta^{\{\mu}\Delta^\nu P^{\rho}P^{\sigma\}}B_{40} - 2\Delta^{\{\mu} \Delta^{\nu} \Delta^{\rho} \Delta^{\sigma\}} C_{40}\big]  
\end{align}
where the GFFs $A_{ij},\,B_{ij},\,C_{ij}$ are a function of $-t$ and are frame independent. Here we indicate the initial momentum as $p$ and the final momentum as $p'$. Also, $P=(p+p')/2$ and $\Delta=p-p'$. $K$ is a kinematic factor dependent on the normalization on the meson state, defined as $K=\frac{1}{\sqrt{4E(p)E(p')}}$ with $E(p)=\sqrt{m_M^2+\vec{p}}$ dependent on the meson mass $m_M$ and the momentum $\vec{p}$. In these proceedings, we focus on the GFFs corresponding to the first-order derivative vector operators in the same fashion as we have done for the scalar, vector, and tensor form factors in Refs.~\cite{Alexandrou:2021ztx,Alexandrou:2021cns}. We also update the work of Refs.~\cite{Alexandrou:2020gxs,Alexandrou:2021mmi} to include the Mellin moments of the pion and kaon PDF up to $\langle x^3 \rangle$.

\section{Results}

\subsection{Mellin moments of PDFs}

Our calculations include the forward limit of the FFs and GFFs ($-t=0$), allowing us to extract the Mellin moments of the PDFs up to the third-order derivative operator. Results for the ensemble cA211.30.32 (at lower statistics) and a detailed methodology for extraction have previously been published in Refs.~\cite{Alexandrou:2021ejy,Alexandrou:2020gxs,Alexandrou:2021mmi}. In Tab.~\ref{tab:pion_moments} and Tab.~\ref{tab:kaon_moments}, we present updated results for ensemble cA211.30.32 and preliminary results for ensemble cB211.25.48 for $\langle x^n \rangle$ with $n=1,2,3$. The higher two Mellin moments, $n=2,3$, are calculated using operators that avoid both finite and power-divergent mixing under renormalization. Also, we extract the multiplicative renormalization functions non-perturbatively~\cite{Alexandrou:2021ejy,Alexandrou:2020gxs,Alexandrou:2021mmi}. For both cases we use results obtained from a two-state fit. Comparing the numerical values in Tables.~\ref{tab:pion_moments} - \ref{tab:kaon_moments}, we find that the two ensembles produce results displaying some differences, likely due to discretization and volume effects. We note that the statistics in the second ensemble is much lower (see Table~\ref{tab:lattice_params}). The differences in the case of the kaon are much smaller, suggesting that the higher mass of the meson reduces some effects, but it is not possible to directly attribute this to the statistics or differences between the lattice parameters. 

\begin{table}[h!]
    \centering
    \begin{tabular}{| c | c | c | c |}
    \hline
    Ensemble  & $\langle x \rangle^{\pi^u}$ & $\langle x^2 \rangle^{\pi^u}$  & $\langle x^3 \rangle^{\pi^u}$ \\
    \hline
  cA211.30.32 & 0.273(07) & 0.112(06) & 0.036(13)\\
  cB211.25.48 & 0.227(14) & 0.090(12) & - \\
    \hline
    \end{tabular}
    \caption{The Mellin moments of the pion for the two ensembles used in this work.}
    \label{tab:pion_moments}
\end{table}

\begin{table}[h!]
    \centering
    \begin{tabular}{| c | c | c | c | c | c | c |}
    \hline
    Ensemble  & $\langle x \rangle^{K^u}$ & $\langle x^2 \rangle^{K^u}$  & $\langle x^3 \rangle^{K^u}$ & $\langle x \rangle^{K^s}$ & $\langle x^2 \rangle^{K^s}$  & $\langle x^3 \rangle^{K^s}$ \\
    \hline
  cA211.30.32 & 0.243(05) & 0.096(2) & 0.042(6) & 0.307(7) & 0.129(4) & 0.060(12) \\
  cB211.25.48 & 0.234(6) & 0.089(4) & 0.060(11) & 0.320(4) & 0.138(2) & 0.082(4) \\
    \hline
    \end{tabular}
    \caption{The Mellin moments of the kaon for the two ensembles used in this work.}
    \label{tab:kaon_moments}
\end{table}

\subsection{Generalized Form Factors}

\begin{figure}[h!]
    \centering
    \includegraphics[scale=0.45]{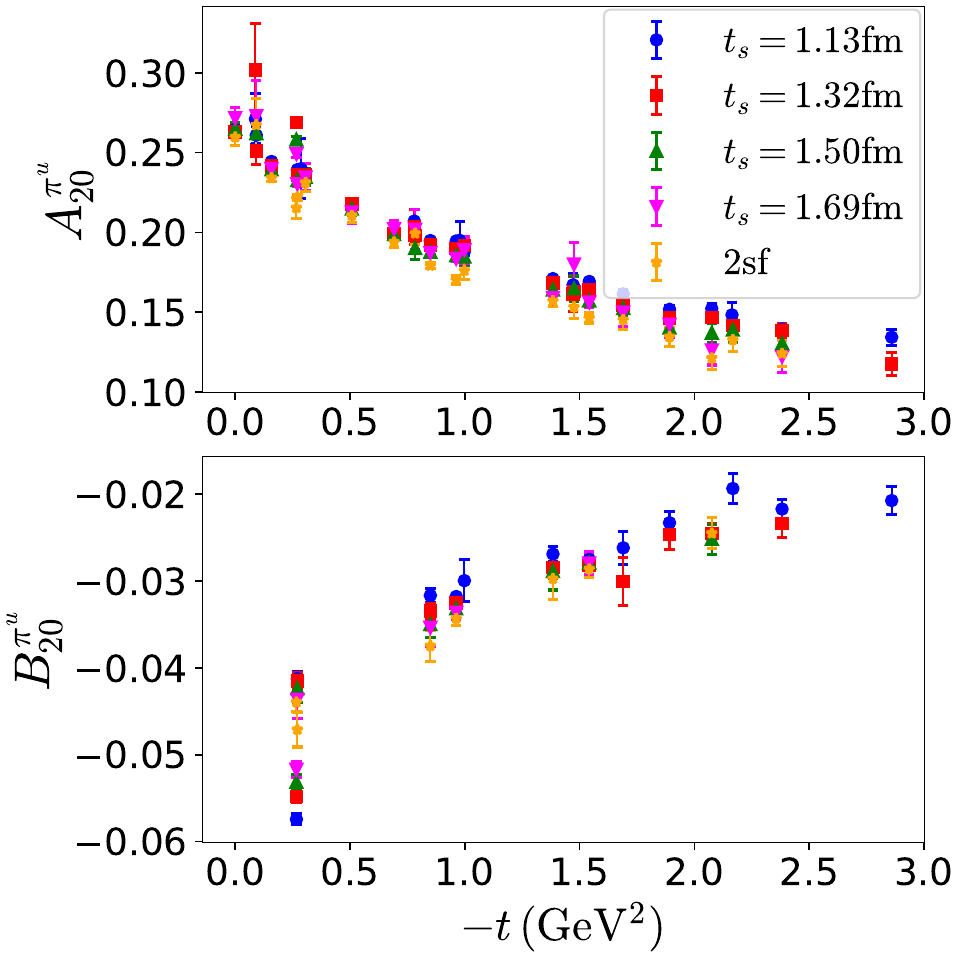}
    \includegraphics[scale=0.45]{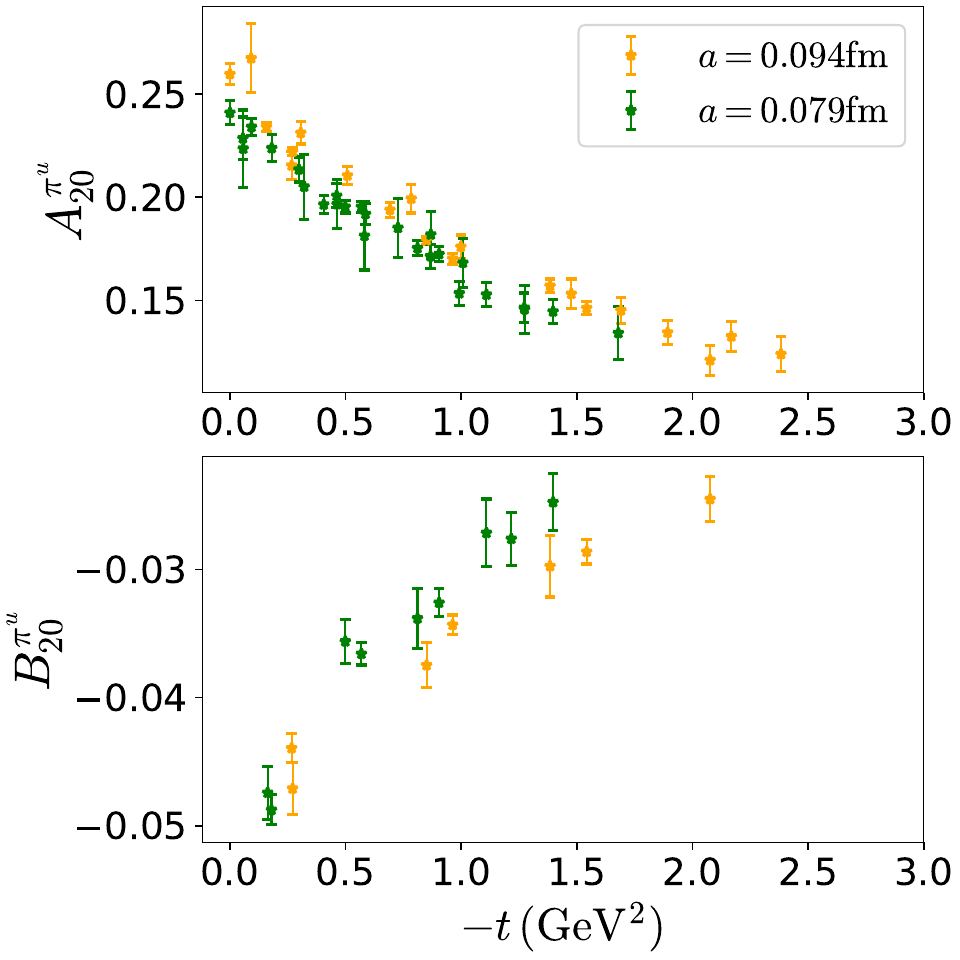}
    \caption{Left: Results for $A^{\pi^u}_{20}$ and $B^{\pi^u}_{20}$ for ensemble cA211.30.32 at source-sink separations $t_s/a=$1.13 (blue circles), 1.32 (red squares), 1.50 (green up-triangles), and 1.69 fm (magenta down-triangles) as well as two-state fit results (orange stars). Right: Comparison of two-state fit results for $A^{\pi^u}_{20}$ and $B^{\pi^u}_{20}$ between ensemble cA211.30.32 (orange stars) and cB211.25.48 (green stars) labelled by their lattice spacings.}
    \label{fig:pion_GFFs}
\end{figure}

Here we show the results for the form factors $A_{20}$ and $B_{20}$ extracted from operators with one covariant derivative. In Fig.~\ref{fig:pion_GFFs} we show the results for the pion GFFs. The results from ensemble cA211.30.32 are shown at the four source-sink separations as well as the two-state fit results. As the statistical errors do not increase linearly with $\vec{q}$, we perform a careful analysis to select points with controlled uncertainty. We find good agreement between all source-sink separations and the two-state fit. Comparing the two-state fits between both ensembles, we see that there are more prominent systematic effects in $B_{20}$ and that it is more difficult to extract points with controlled errors. It is also clear that there are some discretization effects, although they appear smaller than previously observed in the scalar, vector, and tensor form factors.

\begin{figure}[h!]
    \centering
    \includegraphics[scale=0.45]{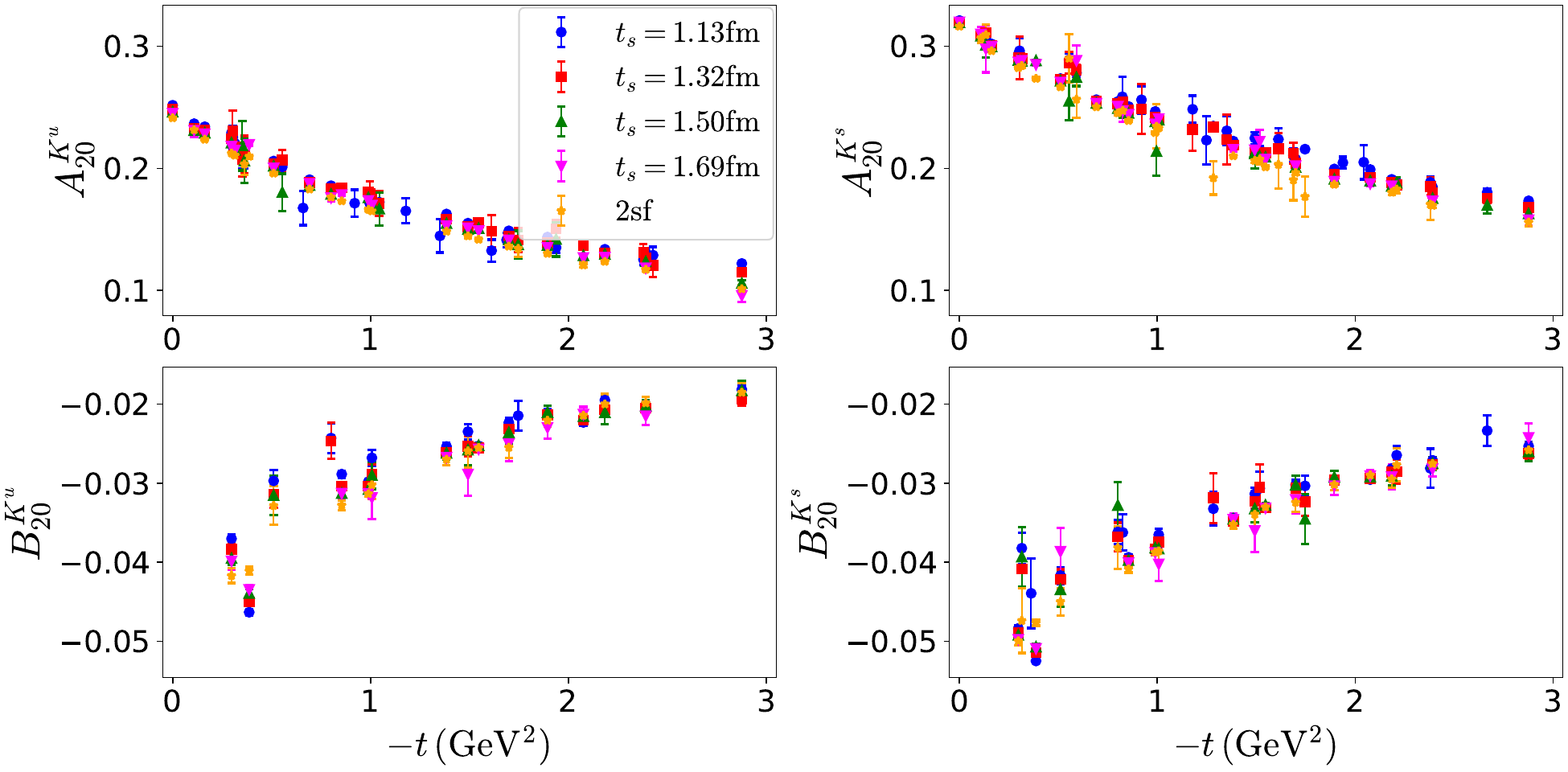}
    \caption{Results for $A^{K}_{20}$ and $B^{K}_{20}$ for the up-quark (left) and strange-quark (right) from ensemble cA211.30.32. Data markers are the same as Fig.~\ref{fig:pion_GFFs}.}
    \label{fig:A_kaon_GFFs}
\end{figure}

\begin{figure}[h!]
    \centering
    \includegraphics[scale=0.45]{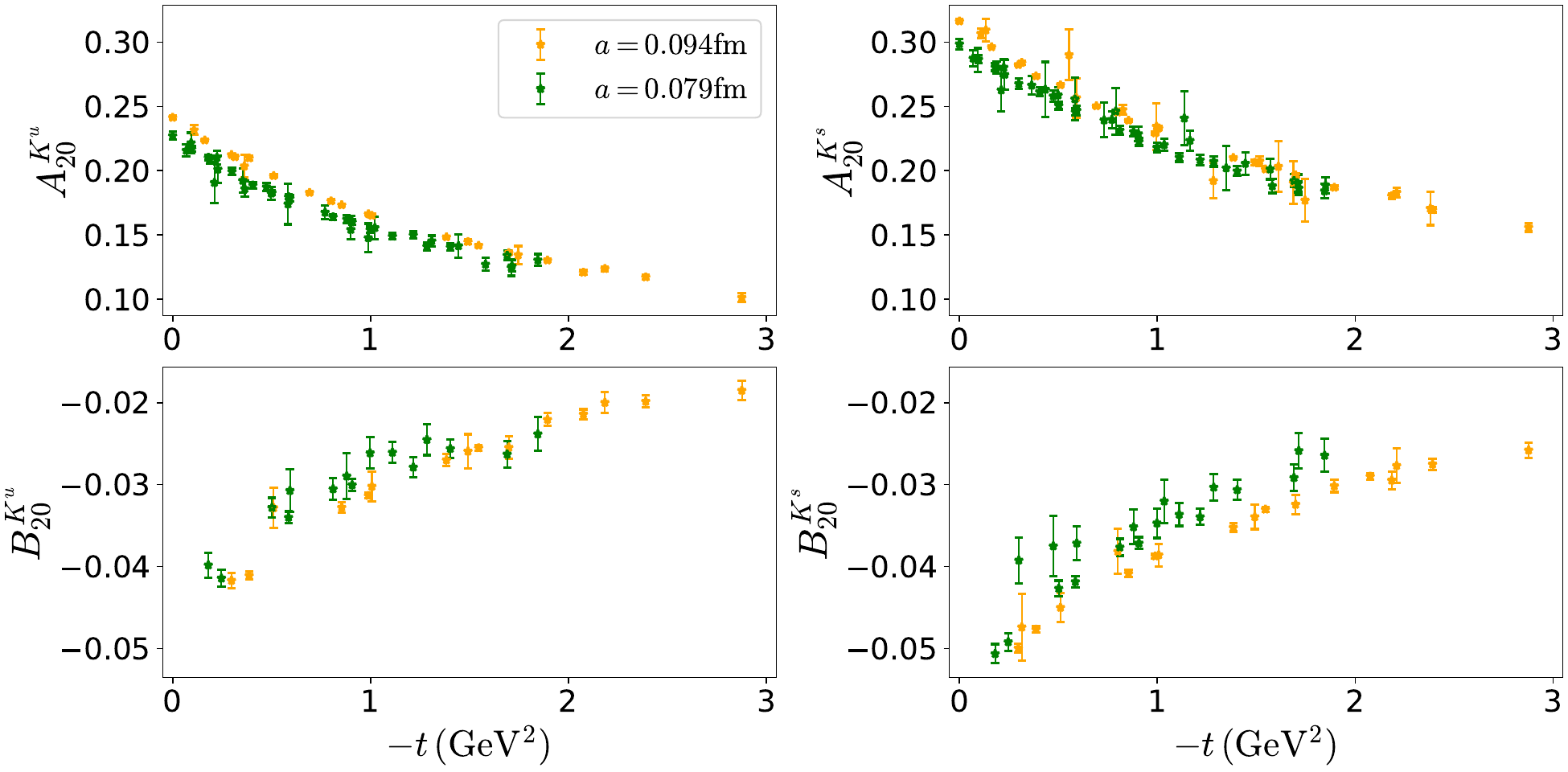}
    \caption{Comparison of $A^K_{20}$ and $B^K_{20}$ for the up-quark and strange-quark between ensemble cA211.30.32 (orange stars) and cB211.25.48 (green stars).}
    \label{fig:kaon_GFF_comp}
\end{figure}

In Fig.~\ref{fig:A_kaon_GFFs} we show the results for the up- and strange-quark components of the kaon as above. The kaon's heavier mass reduces the statistical noise compared with the pion allowing for extraction of GFF data with controlled errors at more values of $-t$. The strange component is dominant in magnitude compared with the up component. We observe some excited state effects, but these are typically within errors. In Fig.~\ref{fig:kaon_GFF_comp} we show the comparison of the two-state fits of the kaon GFFs for both ensembles. The accuracy of the data reveals systematic effects in both $A_{20}$ and $B_{20}$.
\subsubsection{Parametrization of $-t$ dependence}

\begin{figure}[h!]
    \centering
    \includegraphics[scale=0.45]{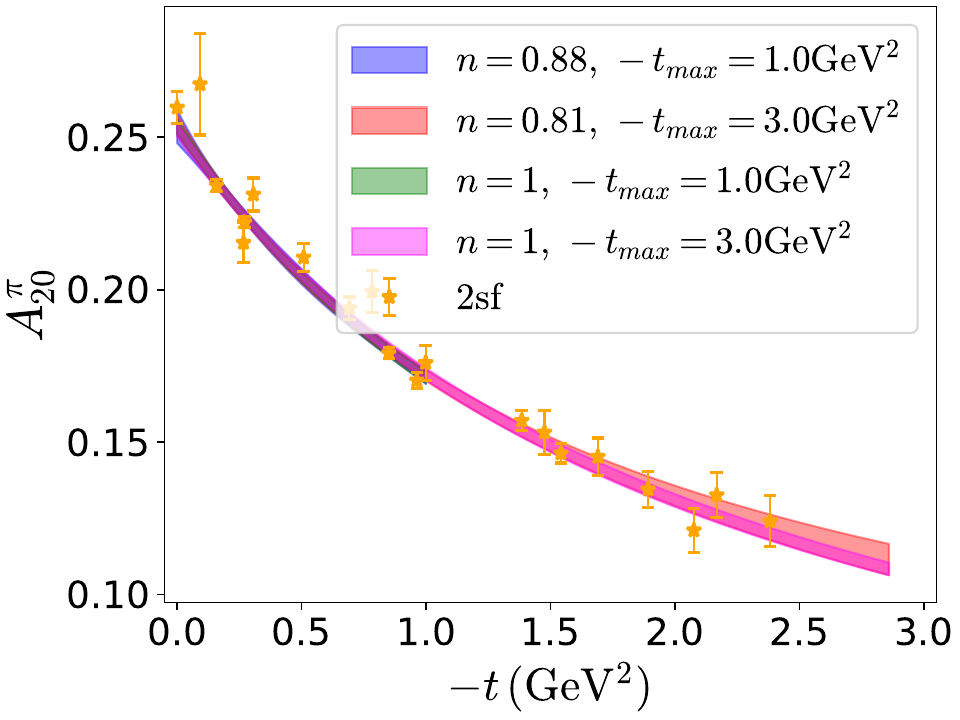}
    \includegraphics[scale=0.45]{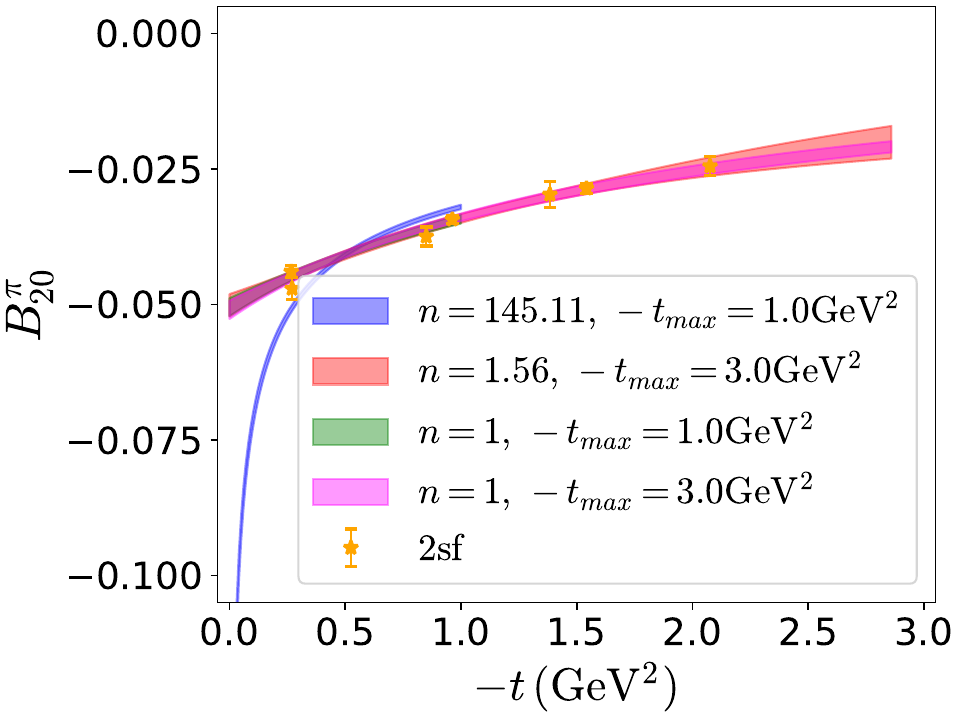}
    \caption{The parametrization of the pion GFF results from ensemble cA211.30.32. Results are shown for the n-pole fit including data up to $-t=1.0$GeV$^2$ (blue) and $-t=3.0$GeV$^2$ (red), and monopole fit results including data up to $-t=1.0$GeV$^2$ (green) and $-t=3.0$GeV$^2$ (magenta). Two-state fit data is shown in orange stars. The values of $n$ from the n-pole fits are shown in the plot legend.}
    \label{fig:Pion_fit}
\end{figure}

We now move the discussion to the presentation of the generalized form factors $A_{20}$ and $B_{20}$. We note that the lattice data for the pion and kaon are extracted at differing values of the 4-vector momentum transfer squared, $-t$, because they have different mass. Thus, in order to compare them, we must parametrize the $t$ dependence of the form factors. We take inspiration from the monopole ansatz depicted by the Vector Meson Dominance (VMD) model~\cite{OConnell:1995fwv}, introducing a free degree $n$ in the denominator like
\begin{equation}
\label{eq:fit}
F_\Gamma(-t) = \frac{F_\Gamma(0)}{\big(1 + \frac{(-t)}{{\cal M}_\Gamma^2}\big)^n} \,,
\end{equation}
where $F_\Gamma(0)$ is the forward limit of the form factor, and ${\cal M}_\Gamma$ is the $n$-pole mass. Such a parametrization allows us to test if the monopole fit, $n=1$, is preferred by our lattice data. 
We utilize the results from the two-state fits to ensure that excited states are eliminated. We apply the fit of Eq.~(\ref{eq:fit}), including data up to a maximum $-t$ of 1 GeV$^2$ and 3GeV$^2$ for both the pion and kaon. In Fig.~\ref{fig:Pion_fit}, we compare these parametrizations for the pion GFFs using the monopole and $n$-pole fits. For the case of $A^\pi_{20}$, all four bands appear to describe the data well and agree with each other. We find $n=0.88$ for the fit, including data up to 1 GeV$^2$ and $n=0.81$ for the fit up to 3 GeV$^2$. As the bands agree and the deviation pole-order from the monopole is fairly small, we conclude that a monopole fit is well justified. The picture is more complicated for the case of $B^\pi_{20}$, likely due to the limited amount of data available for the fits. The $n$-pole fit including data only up to 1 GeV$^2$ produces an asymptotic fit with an unrealistic value for $n$. The monopole fits and the $n$-pole including the full data range again describe the data fairly well. With a pole-order of $n=1.56$, it would be necessary to have more data to make a conclusion regarding the validity of the monopole fit. 

\begin{figure}[h!]
    \centering
    \includegraphics[scale=0.45]{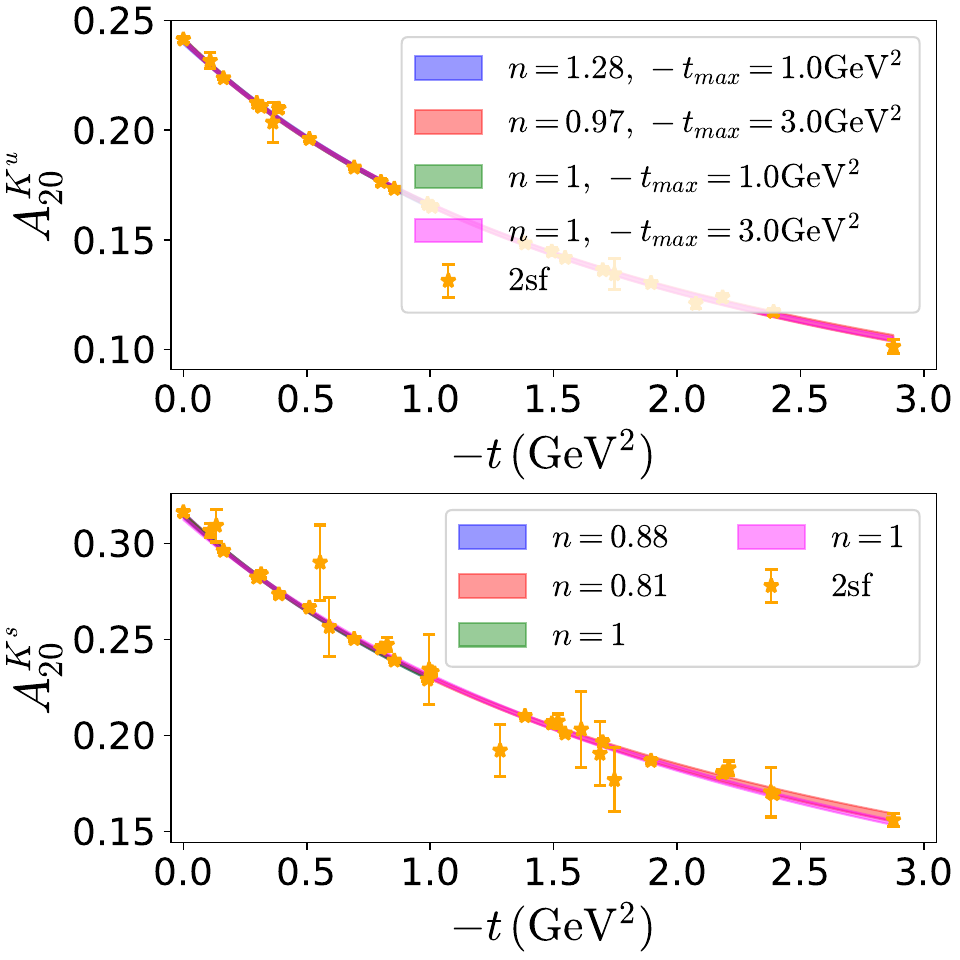}
    \includegraphics[scale=0.45]{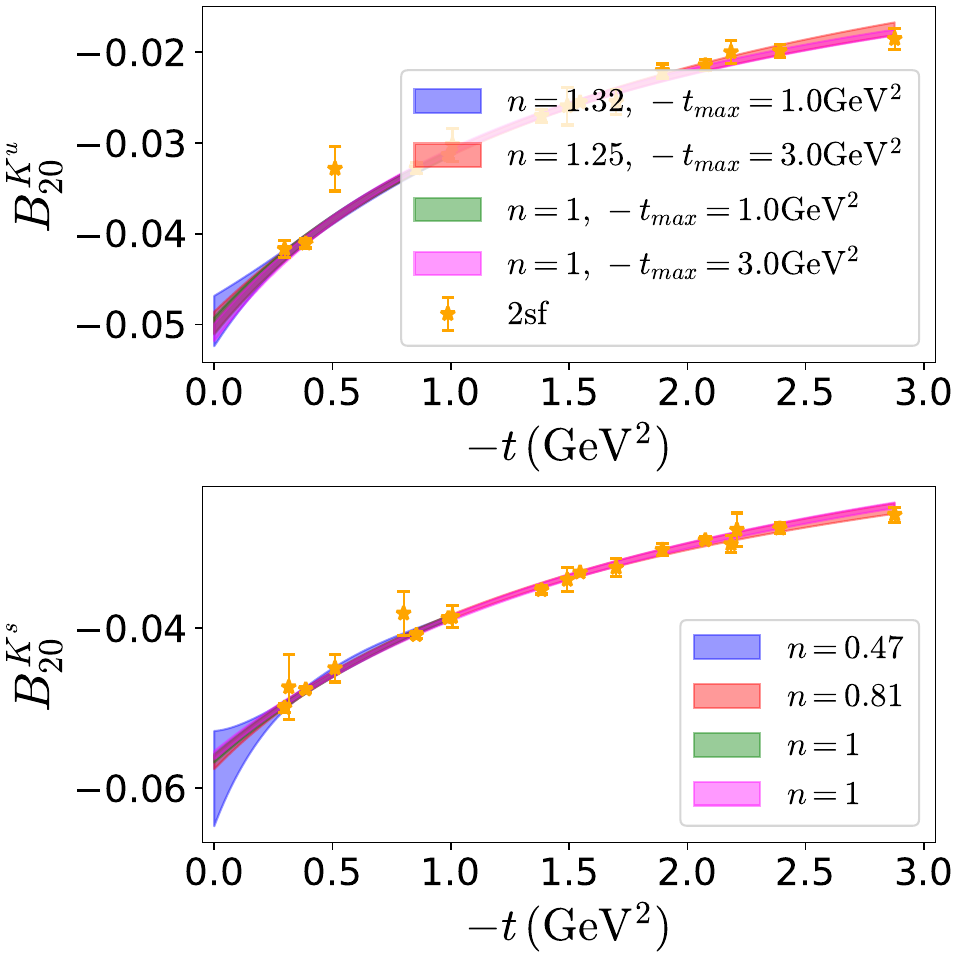}
    \caption{Results for the kaon n-pole and monopole parametrizations. Data format is the same as in Fig.~\ref{fig:Pion_fit}.}
    \label{fig:Kaon_fit}
\end{figure}

We perform the same analysis for the kaon $u$- and $s$-quark contributions, which are shown in Fig.~\ref{fig:Kaon_fit}. In all cases, the data are described well by the fits. Also, all fits agree within errors. Fits on $A^{K^u}_{20}$ and $A^{K^s}_{20}$ reasonably support the use of a monopole fit. For the case of the $B_{20}$ GFF, more data is available for the kaon than for pion case, due to better control of the gauge noise in the heavier kaon. Thus, the various fits are better behaved and indeed describe the data. The pole-order for the full range of data is $n=1.25$ for $B^{K^u}_{20}$ and $n=0.81$ for $B^{K^s}_{20}$. We conclude that it is reasonable to continue with the monopole fit values and the full range of data for the analysis in the next sections. 

\section{SU(3) flavor symmetry breaking}

\begin{figure}[h!]
    \centering
    \includegraphics[scale=0.45]{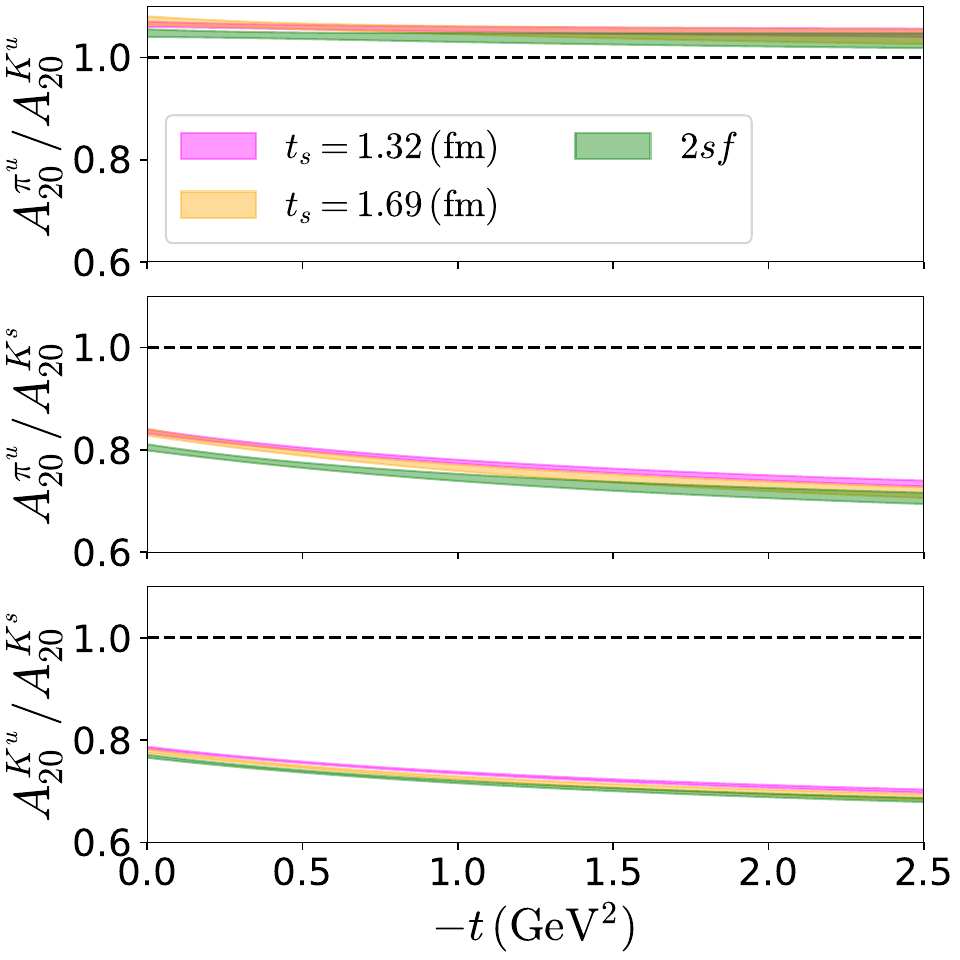}
    \includegraphics[scale=0.45]{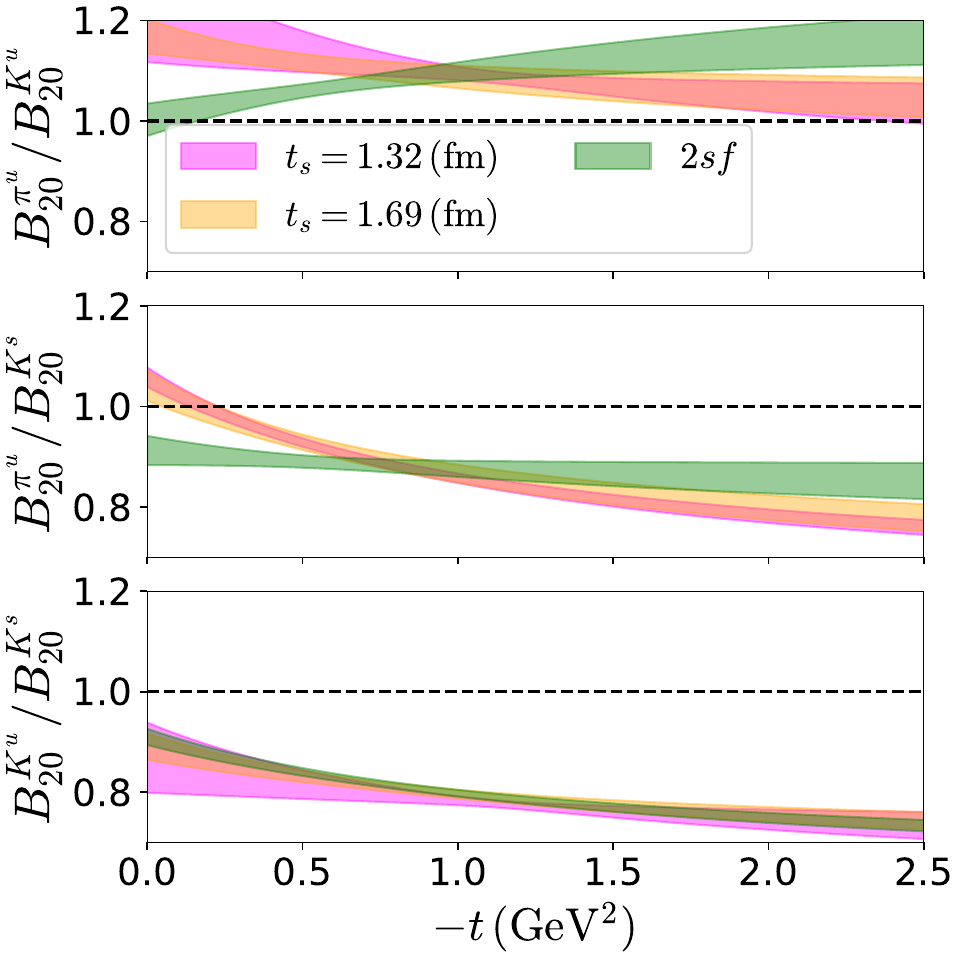}
    \caption{\small{Left: The ratio $A_{20}^{\pi^u}/A_{20}^{K^u}$ (top), $A_{20}^{\pi^u}/A_{20}^{K^s}$ (center), and $A_{20}^{K^u}/A_{20}^{K^s}$ (bottom) for the generalized form factor as a function of $-t$ using the results obtained from the lattice labeled cA211.30.32. Right: The same ratios for the $B_{20}$ GFF. The results for $t_s/a=14,\,18$ and the two-state fit are shown with magenta, yellow, and green bands, respectively.}}
    \label{fig:SU3_old}
\end{figure}

In our previous work~\cite{Alexandrou:2021cns, Alexandrou:2021ztx}, we have observed SU(3) flavor symmetry-breaking effects in the scalar, vector, and tensor form factors by examining the ratios $F^{\pi^u}/F^{K^u}$, $F^{\pi^u}/F^{K^s}$, and $F^{K^u}/F^{K^s}$. This matches what has been observed in nature in the charge radii of $\pi^{\pm}$ and $K^\pm$, as well as in $\pi^{0}$ and $K^0$. Following the analysis of the form factors, we examine SU(3) flavor symmetry-breaking effects in the second Mellin moments of GPDs. Given the difference in the values of $-t$ in the two particles, it is only meaningful to construct such ratios of GFFs using the continuous functions discussed in the previous section. The results are shown in Fig.~\ref{fig:SU3_old} for $A_{20}$ and $B_{20}$. We find significant suppression of excited states compared to the first Mellin moments of GPDs. In the case of $A_{20}$, the ratio of the up-quarks has very little dependence on $-t$ and is just above 1. In the ratios involving the up- and strange-quarks, we find that whether the up-quark comes from the pion or the kaon is insignificant. In both up-strange ratios, the up-quark contribution begins at about $80\%$ of the strange-quark and reduces to about $70\%$ at higher $-t$ values. This differs slightly from what we have observed with the vector form factor. In that case, the ratio begins around 1 and reduces to about $80\%$ with increasing $-t$. The case of $B_{20}$ is not clear due to the relatively small value for the GFF and the large uncertainties of the data. Nevertheless, we are only interested in qualitative conclusions at this stage. As can be seen from Fig.~\ref{fig:SU3_old} the ratio of the up-quark contributions ranges between 0 and 20\%. For the two-state fit, the effects increases with $-t$, while for $t_s=1.32,\,1.69$ fm, it begins at 20\% and decreases with $-t$ increase. SU(3) flavor symmetry-breaking effects are also observed in the ratio of the up-quark in each meson with the strange-quark contribution in the kaon. The effect is between 10 - 30\% with a monotonic behavior as $t_s$ varies.

\begin{figure}[h!]
    \centering
    \includegraphics[scale=0.45]{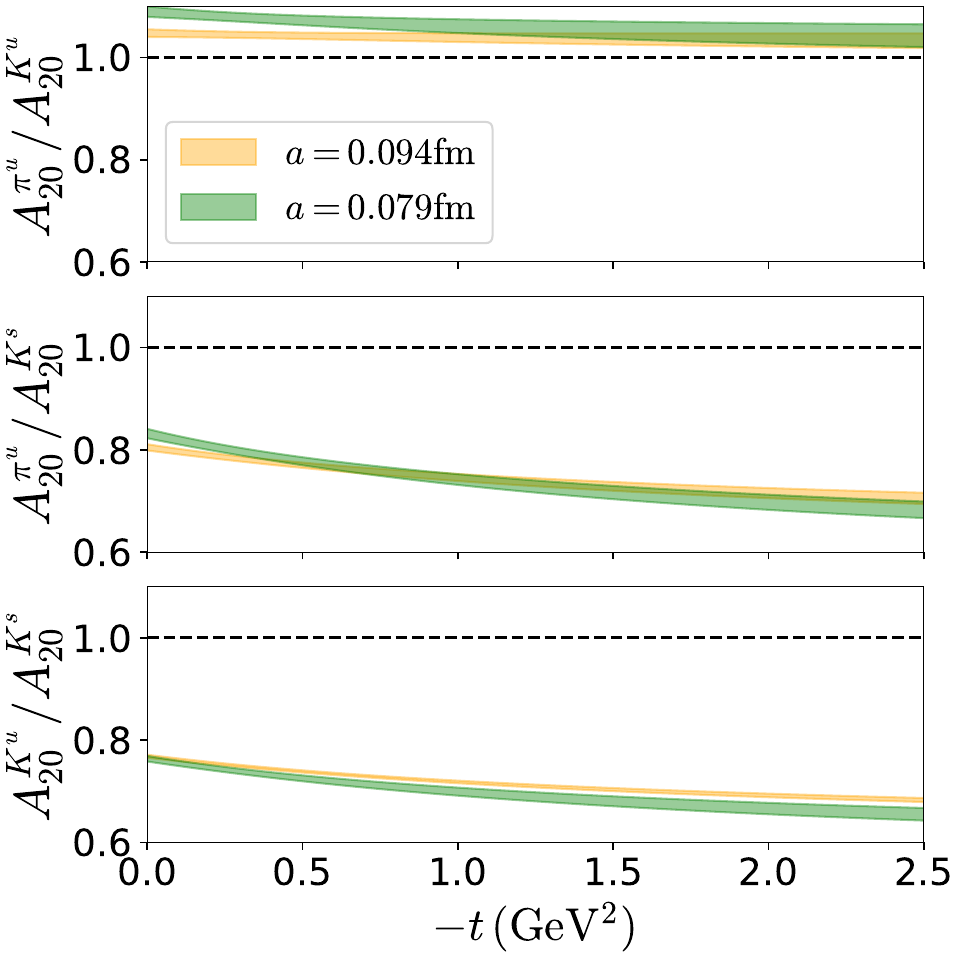}\,\,\,
        \includegraphics[scale=0.45]{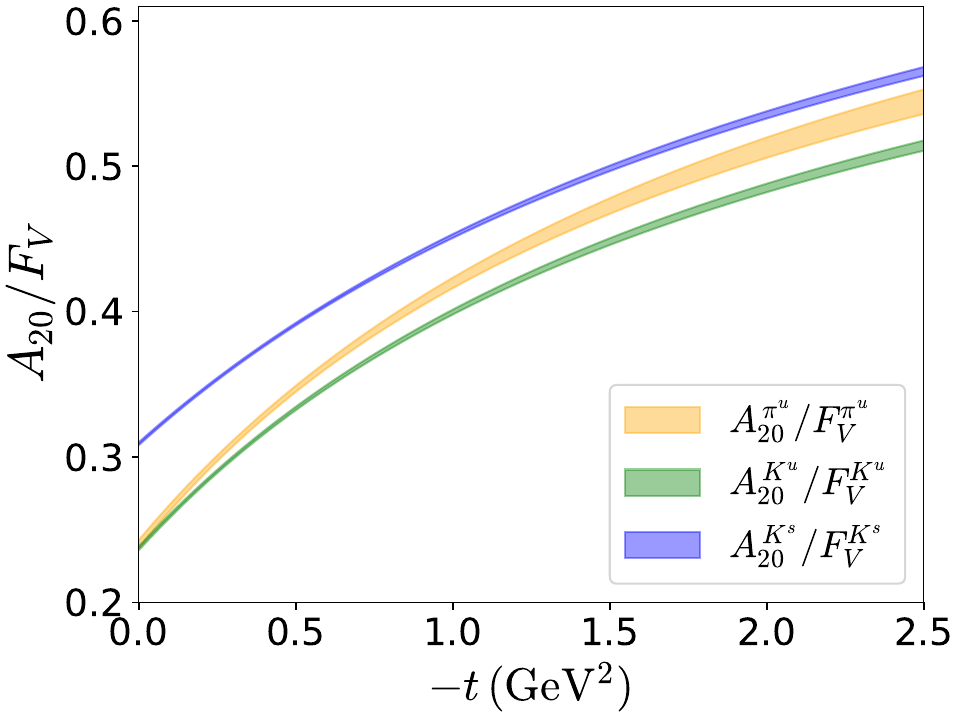}
    \caption{\small{Left: Comparison of the $A_{20}$ ratios for the two ensembles. The ensemble cA211.30.32 with lattice spacing $a=0.094$ fm is shown in yellow, and the ensemble cB211.25.48 with spacing $a=0.079$ fm is shown in green. Right: Ratios of the GFFs $A_{20}^{\pi^u}$, $A_{20}^{K^u}$, and $A_{20}^{K^s}$ with their vector form factor counterparts (yellow, green, and blue, respectively). All ratios are of data from two-state fits.}}
    \label{fig:SU3_comp_hierarchy}
\end{figure}

An interesting aspect is to compare the SU(3) flavor symmetry-breaking effects for the two ensembles we analyzed. Fig.~\ref{fig:SU3_comp_hierarchy} shows a comparison of the ratios as above using the two-state fits of $A_{20}$ for each ensemble. As seen before, the ensemble with finer lattice spacing displays larger statistical uncertainties attributable to the lower available statistics. Even so, the ratios show very similar behavior between the two ensembles in all three cases. 

\subsection{Hierarchy of Mellin moments}

Another component of this work is the comparison of different Mellin moments on the same ensemble, that is, $A_{20}$ and the vector form factor. In the right panel of Fig.~\ref{fig:SU3_comp_hierarchy}, we plot these ratios. The ratios at $-t=0$ are equivalent to the $A_{20}(-t=0)=\langle x \rangle$, as the vector form factor is equal to 1. As the ratios are curves and not flat lines, it is clear that the slopes of the GFFs differ from their vector FF counterparts. We observe that the higher-order Mellin moments are suppressed more slowly than the lower-order ones, and their contributions become about $50\%$ of $F_V$ at $-t=2.5$ GeV$^2$. This is observed across both mesons and flavors.

\section{Summary}

In these proceedings, we present a lattice QCD calculation of the pion and kaon generalized form factors, as well as the Mellin moments $\langle x^n \rangle$ with $n \in [1,3]$. We employ two $N=2{+}1{+}1$ ensembles of twisted mass fermions with clover improvement. The results are renormalized in the $\overline{\mathrm{MS}}$ scheme at 2 GeV. For both ensembles, we utilize a boosted kinematic frame with final momentum $\vec{p}'=\frac{2\pi}{L}(\pm1,\pm1,\pm1)$. We extract the GFFs from matrix elements calculated with three-point functions at four different source-sink separations, as well as data from a two-state fit. The GFF results are self-consistent when comparing source-sink separations and two-state fit results within the same ensemble with limited excited-state effects. When comparing the results between ensembles, it is apparent that the differences between ensemble parameters lead to some systematic effects, in particular discretization and volume effects.

We parametrize the $-t$ dependence of the GFFs using an $n$-pole fit and test the validity of the monopole by setting the pole-order as a free parameter. In most cases, the monopole fit appears well motivated, with any tension attributable to a low number of available data. Using the parametrized data, we examine SU(3) flavor symmetry-breaking effects by taking ratios of the GFFs for the up-quark in the pion and the up- and strange-quark in the kaon. In both ensembles, we observe SU(3) flavor symmetry-breaking effects from 10\% to 30\% when taking ratios of the up and strange contributions. Finally, we examine the relationship between the vector FF and the one-derivative GFF $A_{20}$ finding that the latter decrease more rapidly with increasing $-t$.

\section{Acknowledgements}

We would like to thank all members of ETMC for a very constructive and enjoyable collaboration. 
M.C. and J.D. acknowledge financial support by the U.S. Department of Energy Early Career Award under Grant No.\ DE-SC0020405. S.B. is funded by the project QC4LGT, id number EXCELLENCE/0421/0019, co-financed by the European Regional Development Fund and the Republic of Cyprus through the Research and Innovation Foundation.  S.B. also acknowledges funding from the EuroCC project (grant agreement No. 951740). C.A. acknowledges partial support by the project 3D-nucleon, id number EXCELLENCE/0421/0043, co-financed by the European Regional Development Fund and the Republic of Cyprus through the Research and Innovation Foundation and by the European Joint Doctorate AQTIVATE that received funding from the European Union’s research and innovation program under the Marie Skłodowska-Curie Doctoral Networks action and Grant Agreement No 101072344. I.C. acknowledges support by the U.S.~Department of Energy, Office of Science, Office of Nuclear Physics, contract no.~DE-AC02-06CH11357.
This work used computational resources from Extreme Science and Engineering Discovery Environment (XSEDE), which is supported by National Science Foundation grant number TG-PHY170022. 
It also includes calculations carried out on the HPC resources of Temple University, supported in part by the National Science Foundation through major research instrumentation grant number 1625061 and by the US Army Research Laboratory under contract number W911NF-16-2-0189. 

\bibliographystyle{ieeetr}
\bibliography{references}

%

\end{document}